# Order and randomness in dopant distributions: exploring the thermodynamics of solid solutions from atomically resolved imaging

Lukas Vlcek,[1,2] Shize Yang,[3] Yongji Gong,[4] Pulickel Ajayan,[5] Wu Zhou,[1,*] Matthew F. Chisholm,[6] Maxim Ziatdinov,[6] Rama K. Vasudevan,[6] Sergei V. Kalinin[6]

[1] Materials Science and Technology Division, Oak Ridge National Laboratory,
Oak Ridge TN 37831, USA

[2] Joint Institute for Computational Sciences, University of Tennessee, Knoxville, Oak Ridge, TN 37831, USA

[3] Center for Functional Nanomaterials, Brookhaven National Laboratory, NY

[4] School of Materials Science and Engineering, Beihang University, Beijing 100191, China

[5] Department of Materials Science and NanoEngineering, Rice University,
Houston, Texas 77005, USA

[6] Center for Nanophase Materials Sciences, Oak Ridge National Laboratory,
Oak Ridge TN 37831, USA

[*] Currently at School of Physical Sciences and CAS Center for Excellence in Topological Quantum Computation, University of Chinese Academy of Sciences, Beijing 100049, China.

**Abstract**

Exploration of structure-property relationships as a function of dopant concentration is commonly based on mean field theories for solid solutions. However, such theories that work well for semiconductors tend to fail in materials with strong correlations, either in electronic behavior or chemical segregation. In these cases, the details of atomic arrangements are generally not explored and analyzed. The knowledge of the generative physics and chemistry of the material can obviate this problem, since defect configuration libraries as stochastic representation of atomic level structures can be generated, or parameters of mesoscopic thermodynamic models can be derived. To obtain such information for improved predictions, we use data from atomically resolved microscopic images that visualize complex structural correlations within the system and translate them into statistical mechanical models of structure formation. Given the significant uncertainties about the microscopic aspects of the material's processing history along with the limited number of available images, we combine model optimization techniques with the principles of statistical hypothesis testing. We demonstrate the approach on data from a series of atomically-resolved scanning transmission electron microscopy images of $Mo_xRe_{1-x}S_2$ at varying ratios of Mo/Re stoichiometries, for which we propose an effective interaction model that is then used to generate atomic configurations and make testable predictions at a range of concentrations and formation temperatures.

Condensed matter physics and materials science are both predicated on tuning physical and chemical functionalities via changes in chemical composition. Paradigmatic examples of this approach are the doping of silicon and other semiconductors that underpins virtually all aspects of semiconductor industry and electronics,[1] compositional tuning of oxides that underpin catalysis, energy technologies, and electroceramics,[2-4] alloying of metals, and many others. From a fundamental perspective, most physical studies are performed (and hence functionalities defined) for single crystal solid solutions, a fact which propelled single crystal growth to be a key enabling component of modern research.

The relationship between the atomistic mechanisms of materials doping and emerging functionalities is highly non-trivial. For many materials such as metals and silicon the electron wavefunctions are sufficiently delocalized that the doping effects can be interpreted within effective mean field models, e.g. via the shift of Fermi level or chemical potential of corresponding mobile species. The residual effects of chemical inhomogeneities can then be described via increased scattering rates and reduced mean free paths for electrons and phonons, or effective resistance, whereas exact positions of dopant species are less relevant. Overall, in these cases doping effects are well-described through effective change of bulk material parameters.[5]

This approach however does not hold for materials with higher levels of disorder, giving rise to intriguing physical behaviors such as Anderson localization.[6] The latter is associated with macroscopically disordered ground states resulting in localization of electronic wavefunctions. Similarly, in systems with localized interactions such as strongly correlated materials,[7-11] complex behaviors emerge that are dependent on the strength and directionality of local interactions.[12] Correspondingly, electronic and functional properties will depend not only on average dopant concentrations, but also on the exact configurations of dopant atoms.[13] For phenomena such as phase transformations, including the nucleation and transformation of domains and associated movement of interfaces during the transformation, the details of local atomic arrangements also become important – here, they determine the magnitude of the pinning of the interface, affect the transformation front geometry and account for roughness, and can thus greatly affect other relevant behaviors.[14-15]

Notably, the statistics of atomic configurations of dopant atoms in real space, and hence the effects of the doping on materials behaviors strongly depend on the interactions between the dopant atoms. The effective attractive interactions between the same type of solid solution components can lead to dopant clustering and, above a certain threshold, to segregation of the second phase below the spinodal line. Similarly, repulsive interactions can lead to the formation of additional periodicity on the length scales determined by dopant concentration. These atomic configurations will correspondingly affect the electron, phonon, ferroelectric, or quantum behaviors of the material. The dopants interactions are strongly temperature dependent as determined by the entropic term of free energy. Hence, in realistic materials dopant distributions can be significantly different from the thermodynamic minimum and determined by the preparation history. Furthermore, nanoscale confinement effects can significantly affect even the equilibrium thermodynamics, leading to stabilization of higher-energy phases, emergence of new phases, broadening the regions of solid solution, and other changes.

These considerations necessitate understanding the thermodynamics and effective dopant interactions in real materials. Advances in atomically resolved techniques such as scanning transmission electron microscopy (STEM),[16-18] scanning tunneling microscopy (STM)[19-20] and non-contact AFM (NC-AFM),[21] and atom probe tomography (APT) have allowed insight into atomic configurations on an atom by atom level. However, quantitative information extracted from these observations has been limited, usually because it is difficult to perform appropriate theory-experiment matching at the length scales of both simulation and experiment. Furthermore, while these experiments can in principle produce libraries of possible atomic configurations and structures, the throughput and hence the statistics of these experiments is generally limited.

Here, we analyze the structure of solid solutions from the series of atomically-resolved images in Scanning Transmission Electron Microscopy and infer the microscopic thermodynamic interactions at the formation temperature. This approach allows us to avoid the statistical bottleneck and develop microscopic and thermodynamic generative models for the solid solution formation that can be used to test alternative hypothesis about the formation of the observed structures and provide extrapolations to multiple concentrations and temperatures.

**I. Theoretical background**

Atomically resolved images provide a wealth of information about the interactions and history of the investigated material. In principle, each atom's chemical identity and position within the structure contains a piece of useful information about the system's physics. However, it is not immediately clear what this information may be and how we can use it. We approach this problem, which requires dealing with potentially large and noisy imaging datasets, by applying statistical and machine learning (ML) techniques to develop physically interpretable statistical mechanical models. Specifically, we use model selection and optimization methods that operate on the space of measurement outcomes.

As an illustration relevant for the current task, we consider the solid solution (exemplified here by $Mo_xRe_{1-x}S_2$) extending over $N$ metal atoms on a regular lattice, where $N \sim 400$. Ignoring structural defects, there are $k = 2^N$ possible elementary outcomes corresponding to different atomic configurations, where each can be represented as a unit basis vector in a $k$-dimensional real valued Hilbert space, $H^k$.[22] We note that for classical systems and in the absence of experiment-specific errors (e.g. mis-identified atoms), the space of measurement outcomes is equivalent to the space of the system's coarse-grained states corresponding to all distinguishable lattice configurations of Mo/Re metal atoms. For large sample numbers, the relative frequency of different configurations collected from repeated measurements converges to the probability distribution of the system surface configurations, where each distribution can be represented as a unit vector on the probability space of all possible distributions.

As shown by Wootters for pure quantum states and by Braunstein and Caves for density matrices,[23-24] the angle between probability vectors, typically referred to as *statistical distance*, presents the natural metric for quantifying distinguishability of physical systems. It is defined as,

$$s^2 = \arccos^2\left(\sum_{i=1}^{k} \sqrt{p_i q_i}\right) \quad (1)$$

where $p_i$ and $q_i$ are the probabilities of states $i$ in systems $P$ and $Q$, and the argument represents a scalar product between $k$-dimensional probability vectors. We have recently proposed to use this metric to measure model quality and used it as an optimization loss function that avoids the pitfalls of other commonly used functions, such as the Kullback-Leibler divergence, simple least squares, or various energy and force matching methods for force field optimization.[25]

We have shown earlier that a convenient loss function for $D$ independent datasets in the form of histograms collected from multiple sources, such as images at different conditions, can be written as,[22]

$$S^2 = \frac{1}{n_{Tot}} \sum_{d=1}^{D} n_d s_d^2 \quad , \quad (2)$$

where $s_d^2$ is squared statistical distance for dataset $d$, $n_d$ is the number of samples in dataset $d$, and $n_{Tot}$ is the total number of samples in all datasets.

The practical challenge in dealing with microscopic imaging data using the outlined formalism is the enormous dimension of the Hilbert space and limited number of samples (individual images), which may often amount to just one. In this situation it is impossible to obtain an accurate estimate of the limiting probability distribution *P* that should be matched by a model. According to the maximum likelihood approach, the probability distribution estimate is equal to the distribution of relative frequencies, which would imply zeros for nearly all states.[26] Consequently, there is virtually no chance of a model matching the particular observed configuration.

An alternative estimate of *P* more suited for dealing with zero counts is to use a non-informative Jeffreys prior over the states, which is a uniform distribution on the probability space and whose effect is equivalent to assigning an extra ½ of a sample to each state. The estimate of the system's probability distribution *P* is then,[26]

$$p_i = \frac{x_i + 1/2}{n + k/2} \qquad (3)$$

where $p_i$ is the estimated probability of state *i* of a *k*-state system, $x_i$ is the number of counts in the histogram bin corresponding to *i*, and $n = \sum_{i=1}^{k} x_i$ is the total number of samples. It is easy to see that in the case of large *k* and small *n* the estimated *P* will be nearly uniform for any measurement, and the optimal model will be therefore random with not enough data to support a more complex model.

To overcome this obstacle and obtain more discriminative information from an image, we can first consider the crystalline system as composed of a large number *m* of subsystems, each with *l* dimensions, $l \ll k$. The original Hilbert space can be then expressed as a direct product of the subsystem spaces, $H^k = H^{l^m} = \bigotimes_{i=1}^{m} H_i^l$. In case the subsystems are uncorrelated because of their spatial separation, a lower-dimensional space obtained as the direct sum of subsystem spaces can be formed, $H^{m \times l} = \bigoplus_{i=1}^{m} H_i^l$, which can represent the full physically relevant information. If we further assume that the subsystems are statistically identical as a result of translational symmetry, we can collect all relevant statistics in a single *l*-dimensional space spanning only the states of the subsystem. For a single image we obtain larger number of samples, equal to the number *m* of subsystems, and lower dimension *l* of the subsystem state space. Maximum likelihood or Eqn. (3) will therefore provide a much more accurate estimate of the limiting probabilities that still captures the full relevant information.

We note that this approach is equivalent to the presence of translational statistical invariance in the system and assumes the absence of long-range fields (such as depolarizing field in ferroelectrics). A similar approach was used in the statistical analysis of structural and electronic order parameters using sliding transforms, as reported by Vasudevan et al.[27]

*I.1. Feature selection*

The optimal choice of the subsystems is a feature selection problem. In the limit of large subsystems, we end up with a single sample per image, as discussed above. In the opposite limit of subsystems of the size of a single atom, we can collect a large number of samples, but the two-state (Mo/Re) subsystems will provide only minimal amount of information to discriminate between candidate models because many plausible models can easily fit a binomial distribution (*i.e.*, average concentration). The ideal subsystems that balance the number of samples and the number of distinguishable states $l$ (resolution) will therefore lie in between these extremes and depend on the amount of data. The choice of the most discriminative features will also influence the maximum model complexity that can be supported by the data. As a general rule, when developing models based on microscopic images, we select features that can support the most complex models. Physical considerations of the locality of interactions may guide us to consider features (subsystems) in the form of local configurations that contain information about the direct correlations between atoms that roughly span the range of direct atom-atom interactions.[28] Typically, these may contain the nearest and next-nearest metal atom neighbors (Fig. 2). The statistics of such configurations in the form of histograms represent a natural signature, or fingerprint, of the observed structure, which the model should reproduce. We note that this approach to feature selection is a variation of the bag-of-visual-words ML method used for image classification.[29-31]

*I.2. Statistical hypothesis testing*

Statistical distance, as the geodesic on the probability space, is directly related to the statistical hypothesis testing. In this interpretation, a model of structure formation can be considered a testable hypothesis about the origin of the observed data. While we cannot prove the correctness of the model, we can rule out possible scenarios that are not compatible with the experimental data. For instance, it may not be clear whether configurations observed in microscopic images result from an equilibrium process and can be therefore directly related to interatomic interactions, or whether they represent history-dependent samples from a non-equilibrium distribution.

The target and model distributions of repeated measurement outcomes form multinomial distributions centered around the limiting probability distributions *P* and *Q*, defined on the probability space. In the large sample limit these distributions are well approximated by normal distributions with variance equal to ¼. In this setting, statistical distance can be considered an instance of a Mahalanobis distance *M* defined on the *k-1*-dimensional probability space. We can then use the relation of $M^2$ to p-value,[32] which quantifies the probability that the model generates a distribution that is at least as different as the target distribution. Since $s^2$ follows the $\chi^2_{k-1}$ distribution for *k-1* the degrees of freedom, p-value can be determined as,

$$p = 1 - CDF(\chi^2_{k-1}, 4ns^2) ,  \qquad (4)$$

where *CDF* denotes the cumulative distribution function of $\chi^2_{k-1}$ evaluated at $4ns^2$. Minimizing $s^2$ then results in a model representing a hypothesis that is most difficult to reject using the significance test, *i.e.*, the model distribution is the most difficult to distinguish from the experimental one.

*I.3. Model selection*

As an alternative to the classical statistical significance testing, which evaluates individual models, we can also employ relative model selection criteria. Ideally, we would want to employ the minimum description length (MDL) criterion,[33] which can be interpreted as penalizing model complexity based on the number of distinguishable configurations the model can generate.[34] This criterion is fully consistent with the ideas of the statistical distance framework utilized here. However, for practical reasons we use the simplified version valid in the large sample limit, which coincides with the Bayesian information criterion (BIC),[35] defined here as,

$$BIC = 2ns^2 + \frac{r}{2}\ln n ,  \qquad (5)$$

where the first term is the negative log likelihood of the model generating the observed distribution, *r* is the number of model parameters, and the rest of the symbols have the same meaning as before.

**2. Imaging segregation and phase transition in $Re_xMo_{1-x}S_2$**

As a model system, we have chosen the $Re_xMo_{1-x}S_2$ solid solutions for varying Re concentrations synthesized as described in Materials Section.[36-37] The atomically resolved images across the composition series for x = 0.05, 0.55, 0.78, and 0.95 were acquired on the Nion UltraSTEM100 microscope and are shown in Figure 1. The Re atoms are clearly visible as bright dots, as expected given the higher atomic number of Re.

To analyze the images, we adopt the atom finding algorithm based on the procedure outlined by Somnath et al.[38] Briefly, this involves the first image denoising step via a sliding window reconstruction with principal components, followed by motif-matching and thresholding to find sub-lattices of distinct types and isolate the individual atoms. This functionality is available through the open source python package PyCroscopy.[39-40] Subsequent Gaussian fitting enables sub-pixel accuracy of the atomic coordinates to be determined. Notably, this approach allows not only positional identification of all the atoms in the image, but also classifies them as Mo or Re based on simple thresholding given the change in contrast expected due to higher Z number of Re. The identified atom types are shown superimposed on the atomic contrast in Fig. 1. Thus, obtained data sets contain the information on the atomic configuration of cations in the 2D triangular lattice, i.e. compositional fluctuations. The latter, in turn, can be related to the thermodynamics of the solid solution via the formalism described above.

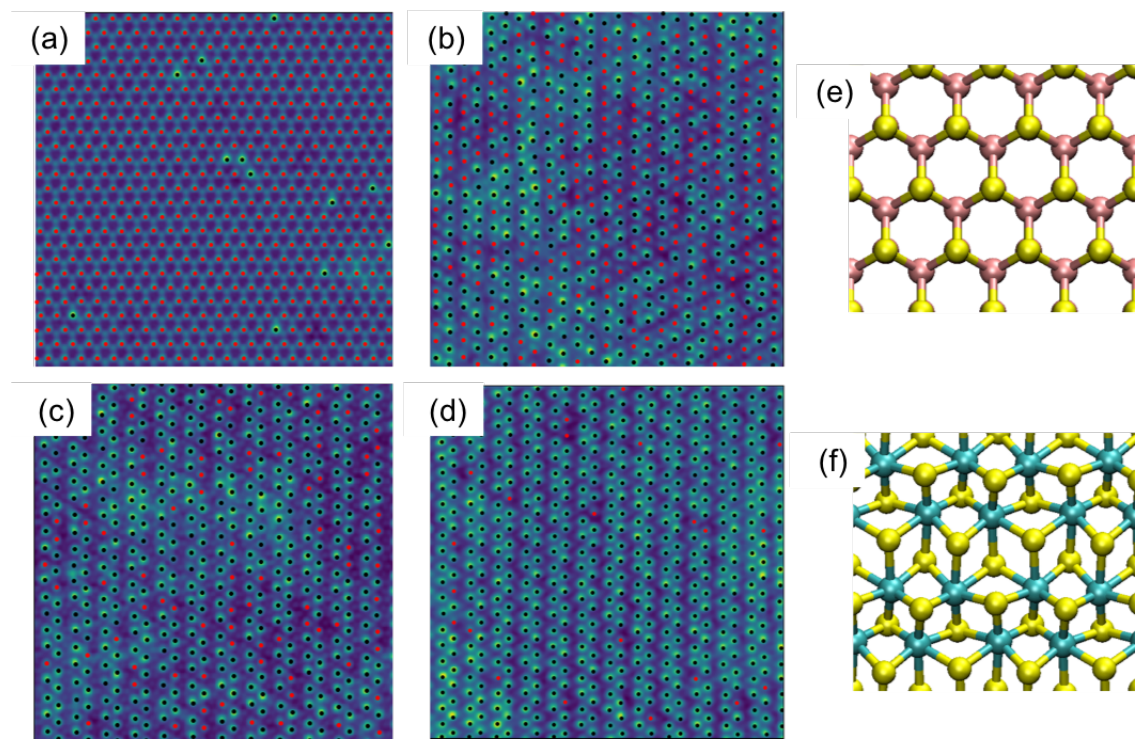

**Figure 1.** STEM images of $Re_xMo_{1-x}S_2$ at different values of $x$: (a) 0.05, (b) 0.55, (c) 0.78, and (d) 0.95. Identified Mo and Re atoms are indicated by red and black dots, respectively. At $x = 0.05$,

the material adopts the MoS$_2$ lattice structure (e), while for higher Re ratios it adopts the ReS$_2$ lattice (f). Color code: Mo (pink), Re (cyan), and S (yellow).

## 2.1. Models of dopant segregation

Here, we restrict our modeling to Mo/Re atom distribution on an idealized hexagonal lattice and ignore defects such as sulfur vacancies. As the first step, we select structural descriptors on this lattice, whose statistics will serve as the target structural fingerprint for model optimization and statistical significance testing. Given the limited amount of data, we constrain our analysis to the statistics of local configurations consisting of an atom and its six nearest neighbors (Fig 2a). Assuming the translational symmetry of the sample, the seven atoms of two possible types can result in $2^7$ configurations. Taking further into account rotational and reflective symmetries, the total number of distinct configurations reduces to 26. The statistics of these configurations in the form of relative frequencies collected from four images at different stoichiometries are shown in Fig. 3.

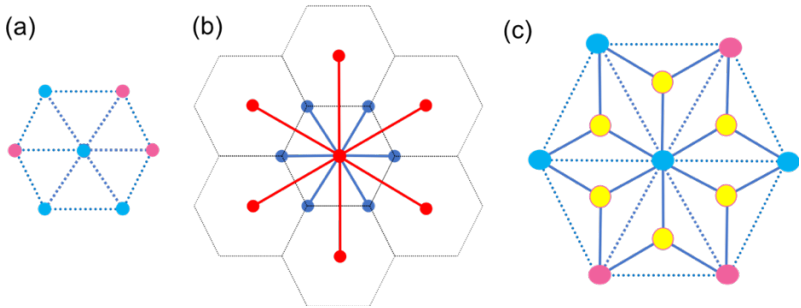

**Figure 2.** (a) An example from a set of 26 local surface configurations whose statistics are to be matched by a model; Mo (pink), Re (cyan). (b) Nearest (blue) and next nearest (red) neighbor metal atom pairs considered in the lattice Hamiltonian of Eq. (6). (c) The triplets of Mo and Re atoms connected to individual sulfur (yellow) atoms define the many body Hamiltonian of Eq. (7).

The complexity of the models reproducing the statistics of local configurations can theoretically range from a null model with zero adjustable parameters and single probability distribution to a model with 4 x 25 parameters, each of which controls the statistics of individual histogram bins collected from the four images. Such a model, which is in effect equivalent to that described by Eq. (3), achieves the maximum complexity with possible probability distributions spanning the entire probability space. Clearly, such a model will overfit and therefore possesses limited predictive power. Physically motivated constraints are thus needed to select a lower-dimensional subspace of possible distributions.

*2.2. Null hypothesis - random model*

As the simplest possible model, the null hypothesis for the observed statistics, we assume that the $Mo_xRe_{1-x}S_2$ configurations collected from the four images are completely random. Physically, such a distribution of metal atoms may result from random deposition of Mo and Re atoms without subsequent thermal equilibration. Alternatively, a random distribution of metal atoms could be formed in an equilibrium system in which the differences in the effective energetics of Mo-Mo, Mo-Re, and Re-Re interactions are very weak.

The random model statistics are compared with the target data in Fig. 3. A quick visual comparison of the two histogram sets suggests that most of the variation in the configuration probabilities can be attributed to their symmetry numbers. To make this comparison more quantitative, we calculated the statistical distances between the target and model distributions and the corresponding p-values for data based on individual images as well as for the combined datasets. The results, summarized in Table I under model R, show that while the random model would pass the significance tests at the typical levels of $\alpha = 0.01$ or 0.05 for the images with very low and very high Re concentrations, we can reject it for the intermediate concentrations, as well as for the combined dataset. It does appear that the distributions are non-random, and detectable ordering happens at the intermediate concentrations. The BIC criterion, Eq. (5), with $r = 0$, attains the value of 73.4.

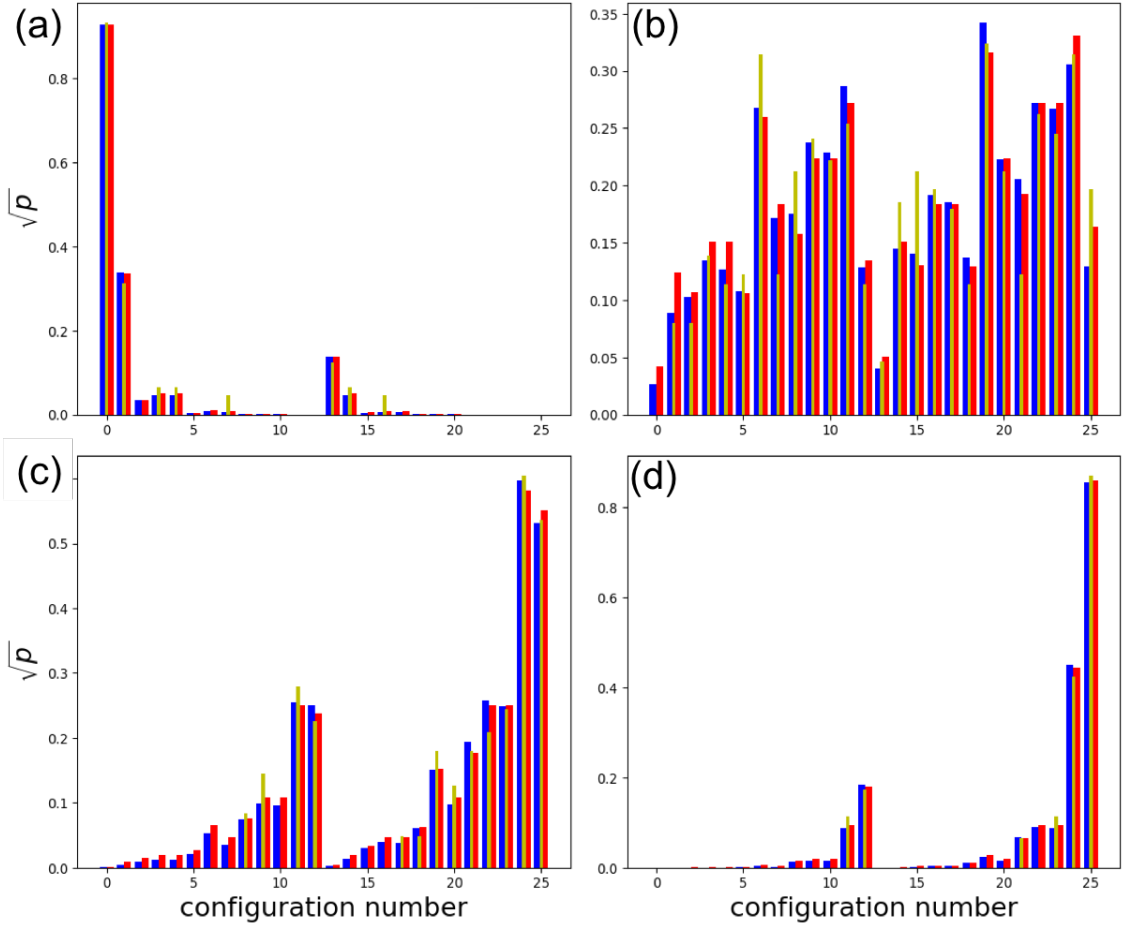

**Figure 3.** Comparison of the square roots of relative frequencies, $\sqrt{p_i}$, of unique local configurations in the target images (yellow), random model (red), and equilibrium model (blue) for 4 compositions studied in the present work. Plots for different values of *x*: (a) 0.05, (b) 0.55, (c) 0.78, and (d) 0.95. The configuration numbers are assigned identification numbers in Supporting Information.

*2.3 Equilibrium pair-additive models*

To probe the segregation hypothesis further, we test a model assuming that the images present equilibrium structures that can be described by a class of models with a simple pair-additive Hamiltonian that includes the first- and second nearest neighbor interactions (Fig. 2b). Both of these interactions effectively account for bonds between Mo and Re atoms mediated by sulfur bridges. The energy of configuration *i* can be written as,

$$u_i = w_1 \sum_{\{NN\}} \delta_{MoRe} + w_2 \sum_{\{NNN\}} \delta_{MoRe}, \qquad (6)$$

where $w_1$ and $w_2$ are interaction energies between Mo and Re atoms in the nearest and next-nearest neighbor positions, respectively; the summation runs over all nearest and next-nearest atom pairs with $\delta_{MoRe} = 1$ a for Mo-Re pairs and $\delta_{MoRe} = 0$ otherwise. This class contains our null hypothesis as a special case with the interaction parameters set to zero, and also a subclass of nearest neighbor models with $w_2 = 0$.

The interaction parameters were optimized to minimize statistical distance between the target histograms and those collected from equilibrium Monte Carlo simulations with the model. As described in Methods section, we combined five reference simulations with tentative models to construct the profile of the combined squared statistical distance $S^2$ as a function of interaction parameters (Fig. 4a). The minimum of this profile was found at $w_1 = -0.1$ and $w_2 = -0.06$. Examples of configurations generated by the equilibrium model at different stoichiometries are presented in Fig. 5. While at the low and high Re ratios $x$ the configurations appear random, ordering of like atoms into smaller clusters seems present at the intermediate concentrations. Even though the profiles of Helmholtz free energy and excess entropy in Fig. 6 indicate increased order at $x \sim 0.5$ (negative excess entropy), they are essentially featureless and do not indicate any phase separation, as can be expected from the attractive effective interactions between Mo and Re atoms (or, equivalently, repulsion between like atoms).

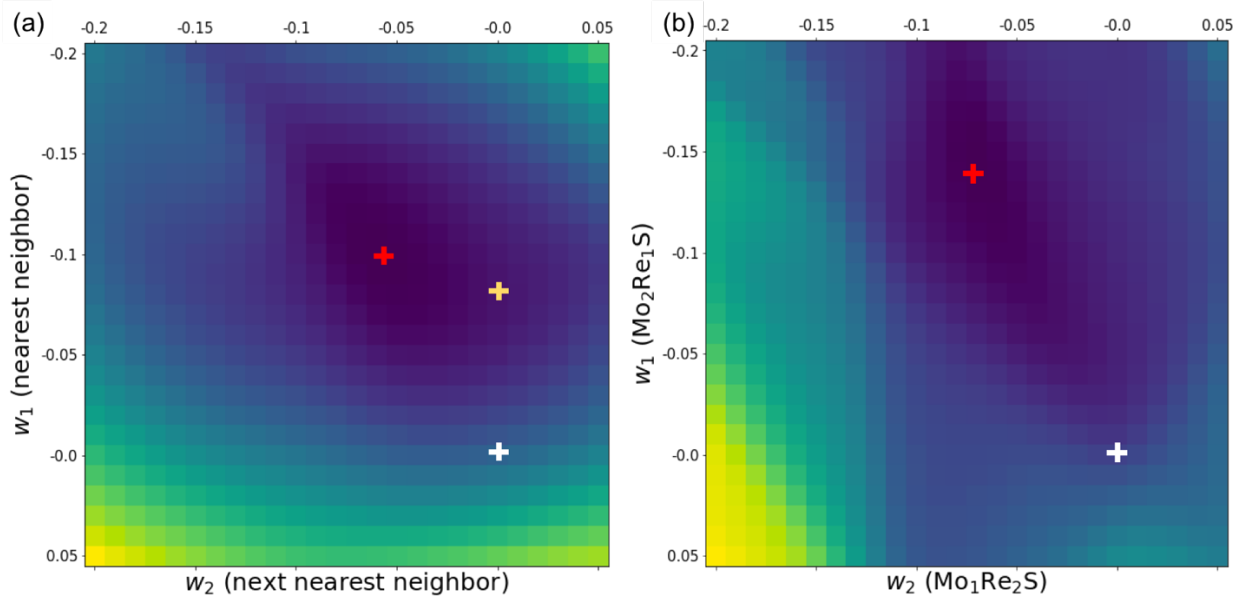

**Figure 4.** Statistical distance profiles for the combined dataset consisting of all four images, Eq. (2), as a function of interaction parameters $w_1$ and $w_2$ of the effective Hamiltonian defined by Eq.

(6) on the left (a), and Eq. (7) on the right (b). Darker colors denote lower values of the loss function, with the minimum for each of the two-parameter models indicated by a red cross, for the single parameter model by a yellow cross, and the random model by a white cross.

To quantify the agreement between these structures and the target images, we performed hypothesis testing. The p-values summarized in Table I (under model P2) show that the equilibrium model is more difficult to reject using standard hypothesis testing. Similar to the random model, it would also pass as a generator of configurations at the two extreme concentrations, $x = 0.05$ and $0.95$, but performs better for the intermediate concentrations with p-values approximately an order of magnitude larger. This improvement means that the model would pass the test for $x = 0.78$ at the significance level of $\alpha = 0.01$. However, it would still fail to explain the configuration statistics at $x = 0.55$. Given the closeness of the random and equilibrium models, we can be also certain that any transition between these two that would represent partial equilibration from the random state would not pass the significance tests. The BIC for the equilibrium model, Eq. (6), with $r = 2$, attains the value of 73.6, which is nearly identical to the random model. Therefore, the improvement in statistical distance (Table I) does not fully justify the two-parameter pair-additive equilibrium model. A simpler pair-additive model can be easily obtained by restricting the interaction parameter to the nearest neighbors by setting $w_2 = 0$ and optimizing only $w_1$. The optimum of $s^2$ is then found at $w_1 = -0.08$, as indicated in Fig 4a. While this choice slightly deteriorates the p-value and $s^2$, the BIC for this lower-complexity model with $r = 1$ is found to be 71.4, which is more favorable than both the two-parameter and random models. Therefore, accepting this criterion, the amount of available data can justify the choice of the simple nearest neighbor model.

*2.4 Equilibrium manybody model*
We may speculate that the overall poor agreement of our pair-additive models stems from their inability to capture the correct form of physical interactions across the range of stoichiometries. In particular, they do not explicitly account for the different bonding topologies of the $MoS_2$ and $ReS_2$ lattices identified at low and high $x$ values, respectively. To test an alternative model of bonding interactions within the system, we constructed a model with a simple manybody Hamiltonian that reflects bonding between triplets of metal atoms sharing the same sulfur atom. Since we are using simulations in the canonical ensemble, which keeps the number of particles of each type constant, we can set the pure-phase energies to zero and only optimize interactions responsible for mixing of Mo-Re atoms. Within this model, the energy of configuration $i$ can be written as,

$$u_i = w_1 \sum_{\{S\}} \delta_{MoMoRe} + w_2 \sum_{\{S\}} \delta_{MoReRe}, \qquad (7)$$

where $w_1$ and $w_2$ are interaction energies of sulfur with $Mo_2Re$ and $MoRe_2$ neighbors; the summation runs over all S atoms with $\delta_{MoMoRe} = 1$ for S with two Mo and one Re bonds, and $\delta_{MoMoRe} = 0$ otherwise; similarly for $\delta_{MoReRe}$ with two Re and one Mo. As in the pair additive model, this model class contains the null hypothesis as a special case with the interaction parameters set to zero.

We followed the same optimization procedure as in the pair-additive model to find the two interaction parameters. The profile of combined squared statistical distance $S^2$ as a function of interaction parameters is shown in Fig. 4b, with the minimum found at $w_1 = -0.14$ and $w_2 = -0.07$. As in the previous cases, the negative interaction coefficients indicate favorable mixing of Mo and Re. The statistical distances and p-values summarized in Table I show that the manybody model is more difficult to reject than the random model based on standard hypothesis testing but performs worse than the simple nearest neighbor model. Taking model complexity into account, the BIC criterion for the equilibrium model, Eq. (6), with $r = 2$, attains the value of 76.1, which is slightly worse than even the random model.

While we were able to find a simple pair-additive model of elemental segregation in $Mo_xRe_{1-x}S_2$, the overall agreement with the imaging data is not completely satisfactory. This indicates that not all physically important effects are captured by the current equilibrium and random models. One possibility to further improve the equilibrium models is to include elastic contributions in the Hamiltonian. A more likely explanation of the discrepancies seems to be the presence of structures created by non-equilibrium processes, whose reproduction would require adequate models. For instance, a model of spinodal decomposition could be tested in a similar manner. However, more data in the form of additional images would be needed to justify selecting a more complex model (equilibrium or dynamic) capable of explaining the observed structures.

Table I: Statistical significance tests and BIC scores of the different models: random (R) and pair additive with one (P1) and two (P2) parameters, pair additive with one parameter (P1), and manybody (M). The columns list the values of sample numbers ($N$), statistical distance ($S^2$), and p-value (PV) for individual and combined datasets.

| dataset | N | $S^2$ (R) | PV (R) | $S^2$ (P1) | PV (P1) | $S^2$ (P2) | PV (P2) | $S^2$ (M) | PV (M) |
|---|---|---|---|---|---|---|---|---|---|
| $x=0.05$ | 464 | 0.0062 | 0.9899 | 0.0067 | 0.9829 | 0.0070 | 0.9771 | 0.0073 | 0.9680 |
| $x=0.55$ | 466 | 0.0325 | 0.0001 | 0.0283 | 0.0009 | 0.0288 | 0.0007 | 0.0284 | 0.0009 |
| $x=0.78$ | 434 | 0.0298 | 0.0013 | 0.0263 | 0.0069 | 0.0247 | 0.0143 | 0.0275 | 0.0040 |
| $x=0.95$ | 471 | 0.0124 | 0.5602 | 0.0129 | 0.4976 | 0.0121 | 0.5852 | 0.0121 | 0.5864 |
| Total | 1835 | 0.0200 | 0.0015 | 0.0184 | 0.0107 | 0.0180 | 0.0168 | 0.0187 | 0.0079 |
| BIC | | 73.4 | | **71.4** | | 73.6 | | 76.1 | |

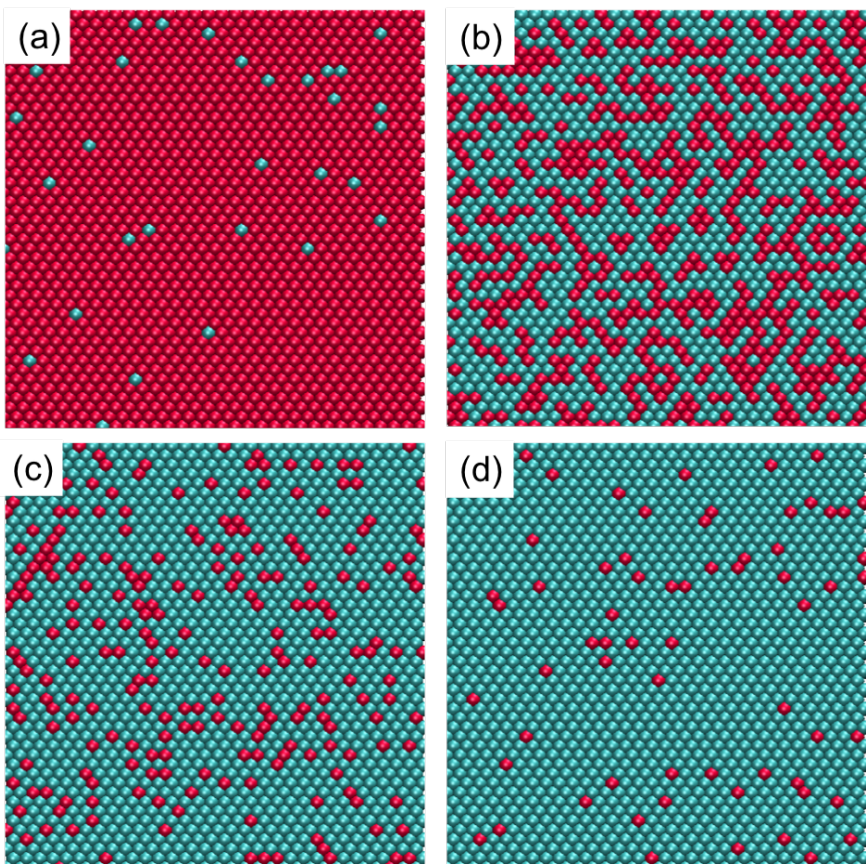

**Figure 5.** Mo (red) and Re (cyan) atom distribution obtained from the equilibrium model for different Re fractions $x$: (a) 0.05, (b) 0.55, (c) 0.78, and (d) 0.95.

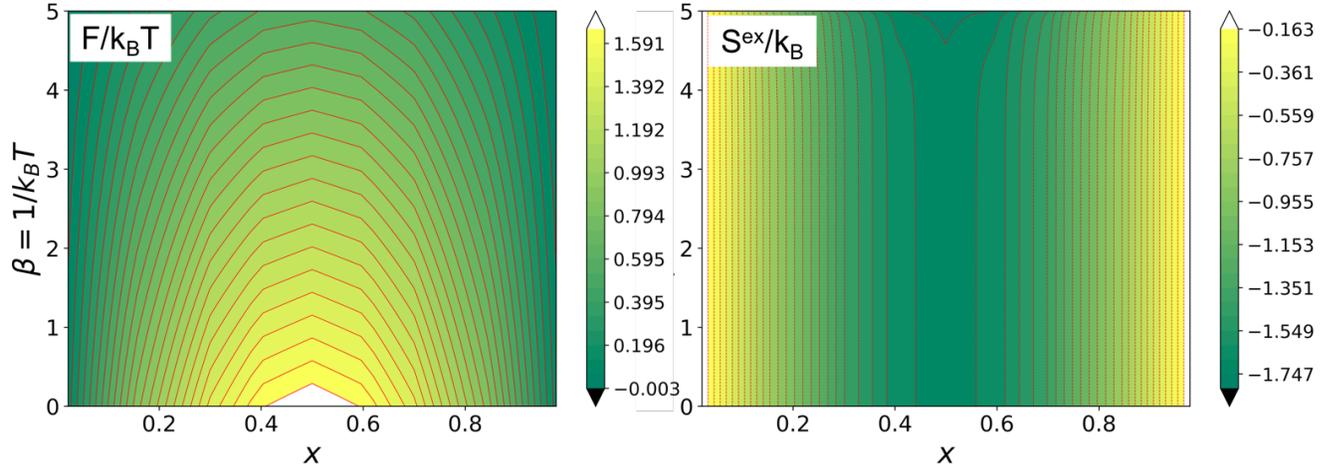

**Figure 6.** Helmholtz free energy $F$ and excess entropy $S^{ex}$ of the equilibrium model as a function of Re fraction $x$ and inverse reduced temperature $\beta$.

## CONCLUSIONS

We have used atomically resolved STEM images of compositional fluctuations in $Mo_xRe_{1-x}S_2$ to develop statistical models of elemental distribution at different stoichiometries. Using these thermodynamic models, we tested alternative hypotheses about the origins of observed structures. While the random model, which ignores any interaction effects (ideal solid solution), appears sufficient to explain the structures observed at low and high ends of the Re relative concentrations, it does not pass the commonly accepted significance levels for the intermediate concentrations, at which the Mo and Re atoms appear to be structured. Alternative equilibrium models with a simple effective pair-additive and manybody Hamiltonians improved the agreement with the observed data but fell short of explaining accurately elemental distributions at comparable concentrations of Re and Mo atoms ($x \sim 0.5$). Based on this analysis, we conclude that the investigated material is close to an ideal solution at forming temperature with weak attractive interactions between the Mo and Re atoms, i.e. tendency for chemical mixing.

We note that while it is difficult to prove that the observed sets of configurations are samples from an equilibrium distribution, and in fact the results indicate that it is unlikely that the structures are not influenced by the material's history, it is possible to test different statistical mechanical models that incorporate both equilibrium and non-equilibrium effects, with the complexity of these models only limited by the amount of available data.

Overall, this approach greatly increases the value of STEM data by allowing it to be connected to the thermodynamic or more complex properties of the system. By the same token, it necessitates the acquisition of much larger data volumes.[41-42] While previously a single image provided qualitative information on the system properties, use of more data enables more statistics, which in turn facilitate improved understanding and discrimination ability between competing models. Furthermore, this approach can be used with data from other experimental tools, including atomic probe tomography, etc., and necessitates the development of automated workflows for data analysis and extraction.

The presented analysis, which integrates statistical mechanics principles with statistical learning methods and statistical hypothesis testing can be easily incorporated into materials science workflows for materials design. In general, the presented work follows the path towards seamless integration of physical theory, machine learning, and experiments. Future work will focus on further development of the unsupervised learning methods for automated feature selection and structure analysis, as well as on expanding the approach to dynamic data and kinetic Monte Carlo modeling.

**Acknowledgements:**

The effort at ORNL including electron microscopy (S.Z.Y., M.F.C., W. Z.) and image analytics and statistical distance minimization (RVK, SVK, LV) was supported by the U.S. Department of Energy, Office of Science, Basic Energy Sciences, Materials Sciences and Engineering Division and performed at the Center for Nanophase Materials Sciences, which is a US DOE Office of Science User Facility.

**Supporting Information**

Supplementary material is available for this manuscript, containing more details about the local configuration statistics.

## METHODS

**Sample growth:**

Molybdenum oxide powder (99%, Sigma Aldrich), sulfur powder (99.5%, Sigma Aldrich) and ammonium perrhenate (99%, Sigma Aldrich) were used as precursors for CVD growth. A selected ratio of molybdenum oxide and ammonium perrhenate was added to an alumina boat with a Si/SiO$_2$ (285 nm) wafer cover. The furnace temperature was ramped to 550 °C in 15 min and then kept at 550 °C for another 15 min for the growth of the Re$_x$Mo$_{1-x}$S$_2$ alloy materials. Sulfur powder in another alumina boat was placed upstream where the temperature was roughly 200 °C. After growth, the furnace was cooled to room temperature using natural convection. The growth process was carried out with 50 SCCM argon at atmospheric pressure.

**Electron microscopy characterization:**

The Re$_x$Mo$_{1-x}$S$_2$ flakes were transferred to TEM grids by spin coating PMMA to support the flakes and etching with KOH to release them from the substrates (by dissolving the SiO$_2$). The annular dark-field images (ADF) were collected using a Nion UltraSTEM100 microscope operated at 60 kV. The as-recorded images were filtered using a Gaussian function (full width half maximum = 0.12 nm) to remove high-frequency noise. The convergence half angle of the electron beam was set to 30 mrad and the collection inner half angle of the ADF detector was 51 mrad. The samples were baked in vacuum at 140 $^0$C overnight before STEM observation. During STEM observation, the probe current was controlled between 10 pA to 60 pA to reduce beam damage.

**Monte Carlo simulations and model optimization**

Simulations with the effective interaction models were performed on a 2-dimensional hexagonal lattice with periodic boundary conditions along the Mo$_x$Re$_{1-x}$S$_2$ plane directions. The simulation cell contained $N=2048$ metal atoms which were equilibrated at reduced temperature $T^* = 1$. After equilibration, the total of $10^5 \times N$ individual MC steps consisting of swaps of Mo and Re atoms were performed in each simulation. The search over the model parameter space to minimize the statistical distance loss function was accomplished with the perturbation technique,[25, 43] which allowed us to minimize the number of MC simulations in the optimization process and reduce thus the computational cost of the inverse problem solution. In the present case of target data with poor statistics, the basic version of the technique based on reweighting the results of a single MC simulation provided inaccurate estimates. Therefore, we used the multistate Bennett acceptance ratio (MBAR) method[44] to combine the results of 5 reference system simulations performed with models with interaction parameters ($w_1$, $w_2$) set to (0, 0), (0.2, 0.0), (-0.2, 0.0), (0.0, 0.2), and (0.0, -2.0).

**SUPPORTING INFORMATION**

**Order and randomness in dopant distributions: exploring the thermodynamics of solid solutions from atomically resolved imaging**

Lukas Vlcek,[1,2] Shize Yang,[3] Yongji Gong,[4] Pulickel Ajayan,[5] Wu Zhou,[1,*] Matthew F. Chisholm,[6] Maxim Ziatdinov,[6] Rama K. Vasudevan,[6] Sergei V. Kalinin[6]

[1] Materials Science and Technology Division, Oak Ridge National Laboratory, Oak Ridge TN 37831, USA

[2] Joint Institute for Computational Sciences, University of Tennessee, Knoxville, Oak Ridge, TN 37831, USA

[3] Center for Functional Nanomaterials, Brookhaven National Laboratory, NY

[4] School of Materials Science and Engineering, Beihang University, Beijing 100191, China

[5] Department of Materials Science and NanoEngineering, Rice University, Houston, Texas 77005, USA

[6] Center for Nanophase Materials Sciences, Oak Ridge National Laboratory, Oak Ridge TN 37831, USA

[*] Currently at School of Physical Sciences and CAS Center for Excellence in Topological Quantum Computation, University of Chinese Academy of Sciences, Beijing 100049, China.

**Identification of configurations in Figure 3.**

Individual local configurations illustrated in Fig. 2 (a), and whose statistics is presented in Fig. 3, can be identified by five-number codes (*i1*, *i2*, *i3*, *i4*, *i5*) as shown in Fig S1. Here *i1* denotes the identity of the central atom (Mo=0, Re=1), *i2* indicates the total number of Re atoms surrounding the central atoms, and *i3*, *i4*, and *i5*, denote the number of Re-Re pairs in the ortho, meta, and para positions, as illustrated in Fig S2.

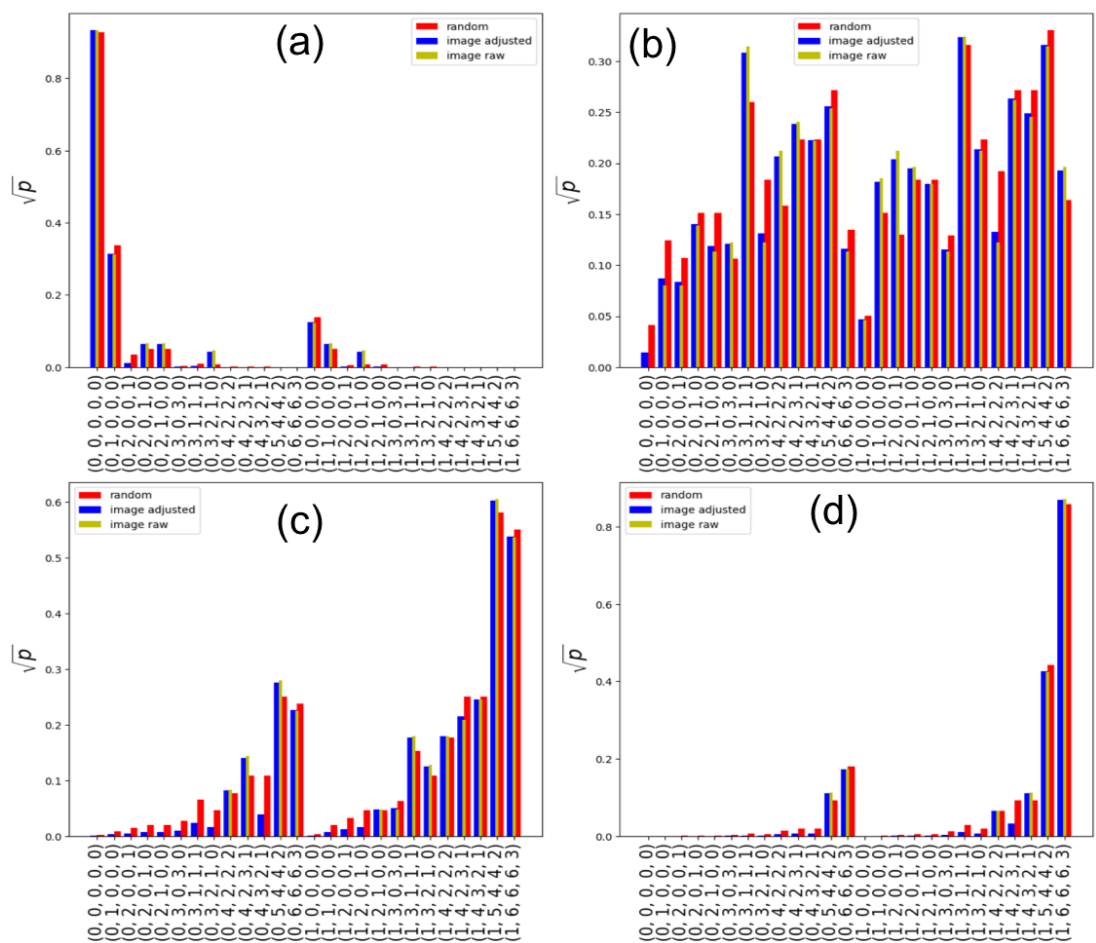

**Figure S1.** Comparison of the square roots of relative frequencies, $\sqrt{p_i}$, of unique local configurations in the target images (yellow), random model (red), and equilibrium model (blue) for 4 compositions studied in the present work. Plots for different values of *x*: (a) 0.05, (b) 0.55, (c) 0.78, and (d) 0.95. Individual configurations are identified by a code illustrated in Fig S1.

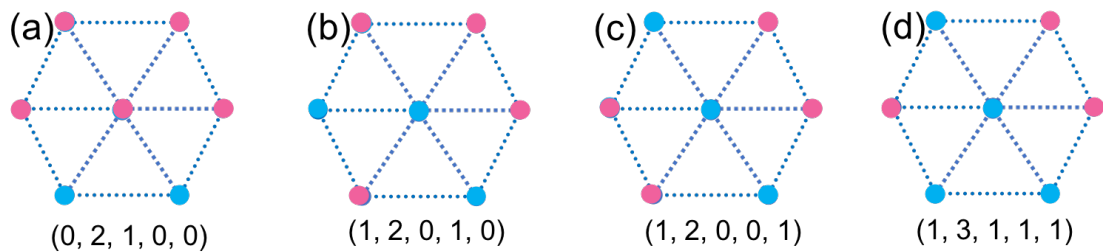

**Figure S2.** Examples of unique local configurations with their identifiers as used in Fig S1; Mo (pink), Re (cyan).